\documentclass[prb,aps]{revtex4}

\usepackage{graphicx}
\usepackage{amssymb,amsmath}

\def \s{\sigma}     
   
\def \th{\theta}

\def \ra{\rightarrow}

\def\lba{\left(}    %don't know if these are already defined, have to check
\def\rba{\right)}
\def\lbc{\left[}
\def\rbc{\right]}
\def\lbb{\left\{}
\def\rbb{\right\}}

\def\be{\begin{equation}}
\def\ee{\end{equation}}
\def \bea{\begin{eqnarray}}
\def \eea{\end{eqnarray}}
\def\bea*{\begin{eqnarray*}}   % for eqnarrays without numbering
\def\eea*{\end{eqnarray*}}

\def \vk{{\bf k}}  % but required with \bf
\def \vr{{\bf r}}  \def \vR{{\bf R}}

\def \multicite{\cite{shikin_many,leiderer_eva,gorkov,ikezi_dimple}}

\begin{document}

\title{Structural Transition of Wigner Crystal on Liquid Substrate}

\author{M. Haque}
\email{masud@physics.rutgers.edu}
\author{I. Paul}
\email{ipaul@physics.rutgers.edu}
\author{S. Pankov}
%\email{pankov@physics.rutgers.edu}  %  ask pankov

\affiliation{Physics Department, Rutgers University, New Jersey, USA}

\date{\today}

%
%     Abstract
%

\begin{abstract}
The physics of an electron solid, held on a cryogenic liquid surface
by a pressing electric field, is examined in a low-density regime
that has not been explored before.  We consider the effect of the
pressing field in distorting the surface at the position of each
electron and hence inducing an attractive force between the electrons.
The system behavior is described in terms of an interplay between the
repulsive Coulomb interaction and the attractive surface-induced
interaction between individual electrons.  For small densities and
large enough pressing fields, we find a parameter regime where a
square lattice is more favorable than the usual triangular lattice; we
map out the first-order transition curve separating the two lattice
geometries at zero temperature.  In addition, our description allows
an alternate static perspective on the charge-density wave instability
of the system, corresponding to the formation of multi-electron
dimples.

\end{abstract}
        %%  find PACS 
    %%    find keywords

\maketitle

\section{Introduction}

The crystallization of charged particles due to Coulomb repulsion,
first predicted by Wigner \cite{wigner}, has been under discussion in
the context of 2D electron systems on cryogenic substrates
\cite{fisher,grimes,shikin_many,eva} and in semiconductor
heterojunctions \cite{semicond}.  Wigner crystallization has also been
studied in three-dimensional trapped-ion systems \cite{ions}.  The
situation in semiconductor junctions, unfortunately, has been obscured
by the presence of disorder \cite{semicond}.  Electrons held on a
liquid helium surface by a pressing electric field, on the other hand,
provided the first clean realization of a Wigner crystal \cite{fisher,
grimes}.  

For the case of electrons on a cryogenic liquid substrate, the
presence of a distortion-prone surface introduces additional physics.
In this paper we investigate a previously-unexplored effect of the
surface on the electron crystal, in a low-density regime that has not
yet been probed experimentally.

Experiments with two-dimensional electron systems on liquid helium are
done with a pressing field $E_\perp$ that holds the electrons to the
liquid surface.  One effect of the pressing field, that has been
studied in depth by a number of authors
\multicite, concerns a regime of density high
enough so that one can treat the surface as a uniformly charged sheet.
In this regime of large electron density $n$, as one increases
$E_\perp$ to a certain value, the surface breaks up into many-electron
dimples, the size of the dimples being given by the capillary length
of the liquid.  The dimples themselves form a triangular lattice
\multicite.  This situation describes a
significant part of the $E_\perp$-$n$ phase diagram.

At smaller densities, it seems sensible to consider the effect of the
pressing field on individual electrons: each electron forms a
single-electron dimple.  (Since the physics determining the shape of
these dimples is the same as that for multi-electron dimples, they
have the same shape, on different scales.)  Single-electron dimples
have been studied previously, especially in the context of forming a
self-trapped polaronic state \cite{polaron}.  However, the formation
of such dimples for each electron should also cause a
'mattress-effect' attraction between any two electrons, and to the
best of our knowledge the effects of this surface-induced interaction
have not been explored.

The surface-mediated attractive interaction between electrons
introduces new physics in the 2D system.  One possibility is one or
more structural phase transitions.  Our analysis shows that, in a
low-$n$, high-$E_\perp$ segment of the $E_\perp$-$n$ phase diagram, a
square lattice is energetically more favorable than the usual
triangular lattice.  We thus prove the existence of at least one
structural transition.  In addition, some of the traditionally-known
surface effects, such as the surface-buckling instability, can be
interpreted from a fresh perspective using the idea of competition
between attractive and repulsive interactions.

Two-dimensional electrons have been studied on several cryogenic
systems \cite{eva}, e.g., surfaces of liquid $^4$He and liquid $^3$He,
at $^4$He-$^3$He interfaces, the interface between solid and
superfluid $^4$He, etc.  
In recent times, experimental efforts have concentrated on increasing
electron densities, and on using thin liquid films, for example for
the purpose of observing a solid-to-liquid quantum phase transition in
the low-temperature, high-density direction \cite{dahm_eva, skachko}.
Other experiments include investigations of the effects of a magnetic
field on excitation and transport properties, and the study and
control of de-coherence \cite{quantum_computing} for quantum-computing
purposes.

In section \ref{section_surfint} we derive the form of the attractive
interaction between two electrons due to surface deformations caused
by the pressing field pushing the electrons down.  In section
\ref{section_genrems} we discuss general properties of systems formed by a
combination of attractive and repulsive forces, and outline the
consequences for the system we are describing.  The energy
calculations for the relative stability of square and triangular
lattices are outlined in section \ref{section_geometries}, and section
\ref{section_0tpd} describes details of the numerics and the resulting
phase-diagram.

\section{Surface-Induced Interaction}  \label{section_surfint}

We consider a system of $N$ electrons, held at positions $\vr_i$ on
the surface of a thick cryogenic (possibly helium) liquid substrate by
a pressing electric field of magnitude $E_\perp$ perpendicular to the
surface.  We will calculate the surface-mediated interaction from
classical, static considerations.

We will use $u(\vr)$ to denote the vertical displacement of the surface
at point $\vr$, as compared to the undistorted (flat) configuration.  We
proceed to write down the energy as a functional of $u(\vr)$.  There
are three contributions: a surface tension term describing the energy
cost due to surface distortion, a pressing-field term describing the
energy that the electron gains due to vertical displacement, and a
gravity term describing the bulk displacement of helium accompanying
the surface distortion.
\[
E[u(\vr)] ~=~ \sigma \int d^2r 
\lbc 1+ \lba \nabla u(\vr)\rba^2 \rbc^{1/2} ~+~
eE_\perp \sum_{i=1}^{N}u(\vr_i) ~+~ \frac{g\rho}{2} \int d^2r \lbc
u(\vr)\rbc^2  \, \, .
\]
Here $\sigma$ and $\rho$ are respectively the surface tension and
density of helium.  The pressing electric field $E_\perp$ actually
contains contributions from both the externally applied field and the
field due to the image charge formed by the helium dielectric.  We
will be interested in large applied fields, compared to which the
dielectric effect is negligible.

Expanding the surface-tension term to lowest order in $\nabla{u}$, we
get in momentum space:
\[
E[u(\vk)]^{\{\vr_1,\ldots,\vr_N\}} ~=~ 
\frac{A\sigma}{2} \sum_\vk k^2 u(\vk) u(-\vk)
~+~ eE_\perp \sum_{i=1}^{N} \sum_\vk u(\vk) e^{i\vk\cdot\vr_i} 
~+~ \frac{Ag\rho}{2} \sum_\vk  u(\vk) u(-\vk)  \, \, .
\]
The $\vk$ are 2D wave-vectors and $A$ is the area; $4\pi^2\sum_\vk
 \leftrightarrow A\int d^2k$.  The form of $u(\vk)$ is now determined by
 minimizing the energy functional.  The result is  
\be  \label{uk}
u(\vk) ~=~  -~\frac{eE_\perp}{A\sigma}
\sum_{i=1}^{N}  \frac{e^{-i\vk\cdot\vr_i}}{k^2+l_0^{-2}} 
~=~  \sum_{i=1}^{N} u_1^{\vr_i}(\vk)  \, \, .
\ee
Here $l_0 = \sqrt{\s/g\rho}$ is the capillary length of the liquid
substrate; it will be the important length scale in all our
considerations.  Also, 
\[
u_1^{\vr_i}(\vk) =  ~-~\frac{eE_\perp}{A\sigma} ~
 \frac{e^{-i\vk\cdot\vr_i}}{k^2+l_0^{-2}} 
\]
is the Fourier transform of the distortion due to a \emph{single}
electron at $\vr_i$, i.e., the shape of a single-electron dimple, as
can be easily verified by minimizing the energy functional for a
single-electron system.  

The energy of the system is the minimum of the functional $E[u(\vk)]$,
and can be now written in terms of the $u_1^{\vr_i}$'s:
\[
E(\vr_1,\ldots,\vr_N) ~=~ \frac{A\sigma}{2}\sum_\vk 
\lba k^2 + l_0^{-2} \rba \sum_{i=1}^{N} u_1^{\vr_i}(\vk)  
\sum_{j=1}^{N} u_1^{\vr_j}(-\vk) 
~+~ eE_\perp \sum_{i,j} \sum_\vk u_1^{\vr_i} e^{i\vk\cdot\vr_j}  ,
\]
which separates into diagonal ($i=j$) and non-diagonal ($i{\neq}j$)
pieces: 
\begin{equation}  \label{E_min}
E(\vr_1,\ldots,\vr_N) ~=~   N E^{(1)} ~+ \sum_{i<j} V(\vr_i -\vr_j)  ,
\end{equation}
with 
\[
E^{(1)} ~=~ -~ \frac{(eE_\perp)^2}{2A\s} \sum_\vk \lba k^2 + l_0^{-2}
\rba^{-1},  \qquad
V(\vr_i -\vr_j) ~=~ -~ \frac{(eE_\perp)^2}{A\s} \sum_\vk
\frac{e^{i\vk\cdot(\vr_j-\vr_i)}}{k^2 + l_0^{-2}} 
\]
The first term of eq (\ref{E_min}) is an extensive quantity
representing the energy of $N$ independent electrons.  The second term
gives the attractive interaction energy between the electrons mediated
by surface deformation.  We have thus obtained the 'mattress'-effect
attractive potential to be
\begin{equation}   \label{surfinducedV}
V(\vr) ~=~  -~ \frac{(eE_\perp)^2}{4\pi^2\sigma} \int d^2k\
\frac{\cos\lba\vk\cdot\vr\rba}{k^2+1/l_o^2} 
~=~ -~ \frac{(eE_\perp)^2}{2\pi\sigma} K_0(r/l_0) .
\end{equation}
Here $K_0$ is the zeroth-order Bessel function of the second type.  It
has the asymptotic behavior $K_0(x) \sim -\log(x)$ for $x
\ra 0$ and $K_0(x) \sim (\pi/2x)^{1/2} e^{-x}$ for $x \ra \infty$.

Two comments are in order concerning this derivation.  First, the
electrons have been treated as point objects, and this leads to
interactions and dimple shapes that diverge logarithmically at small
distances.  This divergence is cured by the fact that the electron
wavefunction has a finite spatial width.  In this work we will not
attempt a calculation of this wavefunction width, but will assume that
this spatial extension is much smaller than the lattice spacings which
we consider.  This assumption is particularly reasonable in the
low-density region, deep in the solid phase, that we focus on.

Second, we have retained only the $\lba{\nabla}u\rba^2$ term in the
surface-tension energy.  This is equivalent to keeping only the
pairwise interaction between electrons and dropping four-electron and
higher terms.  For quantitative calculations, eq \eqref{surfinducedV}
is used only in a very-low-density regime; effects of four-body or
higher terms are expected to be small here.  Our order-of-magnitude
estimates in section \ref{section_genrems} concerning moderate-to-high
densities makes use only of qualitative ideas from the mattress-effect
calculation, because in this regime the effect of higher-order terms
is expected to be important.

\section{Effects of Attractive Interaction: General Remarks}
\label{section_genrems} 

The Wigner lattice on a helium substrate is formed by electrons which
interact via a Coulomb repulsion and a surface-mediated attraction:
\begin{equation}    \label{potential_total} 
V(\vr) ~=~ V_{\rm coul}(\vr) ~+~ V_{\rm surf}(\vr) 
~=~ \frac{e^2}{r} ~-~ \frac{(eE_\perp)^2}{2\pi\sigma}
K_0(r/l_0) .  
\end{equation}
The effect of the surface term has not been considered in detail
before, and we shall proceed to do so in the present paper.

The surface-induced attraction can be tuned by controlling the
pressing electric field $E_\perp$.  At short enough distances, the
$1/r$ function dominates over the logarithmic surface term, and at
large $r$ the Coulomb term again dominates over the exponentially
decreasing attraction.  $V_{\rm surf}$ can compete with $V_{\rm coul}$
only at intermediate distances.  As $E_\perp$ is ramped up, the
distance scale at which $V_{\rm surf}$ first becomes comparable to
$V_{\rm coul}$ is $r\sim l_0$.

In general, when microscopic objects interact via attractive and
repulsive potentials of different ranges, several things can happen
depending on the relative strengths and ranges of the attractive and
repulsive forces.  First, consider the case of repulsive forces alone,
or short-range repulsion coupled with longer range attraction.  This
situation tends to create microscopic-ordered phases, such as Wigner
crystals, vortex lattices and skyrmion lattices
\cite{skyrmion_lattice}.  Second, when the attractive force dominates
at all distances, the system tends to collapse.  One example is what
happens at the higher critical magnetic field $H_{\rm c2}$ of a
Type-II superconductor: the interaction between vortices of the mixed
phase becomes attractive cite{kramer} at $H = H_{\rm c2}$, and this
leads to collapse of the vortex matter so that the system is filled
with the normal electrons of the vortex cores, and superconductivity
is destroyed.  And finally, a combination of short-range attraction
and long-distance repulsion tends to create intermediate-scale order,
or ``clustering''.  Examples are charge density waves in solids and
quantum hall systems.

This approach enables us to view the well-known electrohydrodynamic
(surface-buckling) instability of this system \multicite from a novel
perspective.  For an electron density much larger than $l_0^{-2}$, any
electron has a large number of electrons within a distance less than
the capillary length $l_0$ from itself, so that the attractive force
due to surface distortion acts between a large number of particles.
Thus we have an attractive interaction at intermediate distances and a
long-distance Coulomb repulsion.  Therefore when the attractive
interaction is ramped up by increasing the electric field, one can
expect from the preceding general discussion the formation of
intermediate-scale clusters, which are themselves ordered in a regular
pattern.

This is exactly the phenomenon of formation of many-electron dimples
that is observed for high-density electrons under large pressing
fields.  Previously this instability has been studied in terms of the
excitation spectrum of a surface approximated as being uniformly
charged \multicite.  In the traditional analysis, one finds that at a
certain pressing field, the spectrum goes ``soft'' at wavenumber
$k\sim l_0^{-1}$, indicating the onset of a charge-density instability
of this wavenumber.  The pressing field at which this spectrum
softening first happens is given by $E_\perp^2 = 4\pi[\rho{g}\s]^{1/2}
- (2{\pi}ne)^2$, or $E_\perp \approx [16\pi^2\rho{g}\s]^{1/4}$ at low
densities.

In our picture of competition between attractive and repulsive forces,
the formation of multi-electron dimples (intermediate-scale order)
would occur when the attractive term $V_{\rm surf}$ starts to dominate
over the repulsive term at intermediate or small distances.  This
viewpoint allows a simple calculation of the pressing field at which
the surface-buckling occurs: it is the pressing field for which we
have $V_{\rm surf}(r=l_0) \approx V_{\rm coul}(r=l_0)$, i.e., $E_\perp
\approx \{\lbc{K_0}(1)\rbc^{-2} 4\pi^2\rho{g}\s\}^{1/4} \approx 1.1
\lbc 16 \pi^2\rho{g}\s\rbc^{1/4}$, within 10\% of the traditional
result.

Next we want to concentrate on a lower-density regime, $n \gtrsim
1/l_0^2$, where we find a less dramatic but nevertheless important
effect of the competition between attractive and repulsive forces.  As
the inter-particle distance approaches $l_0$, the formation of
many-particle dimple becomes less likely, since there is no longer
``many'' electrons within the capillary-length scale, and the
uniform-smeared charge approximation becomes untenable.  It is not
clear whether a surface-buckling instability is still present.
However, one still expects some effects of increasing attraction as
$E_\perp$ is ramped up.  One possibility is a structural transition to
a different lattice geometry.  In the remainder of the paper, we show
that there is indeed a region of $E_\perp$-$n$ space where a square
lattice becomes more favorable to the triangular lattice for the
Wigner crystal.

Bonsall and Maradudin's classic work \cite{bonsall} (referred to as BM
from now on) has shown for a 2D
electron lattice, where the electrons interact via a Coulomb force
only, that the triangular (hexagonal) lattice is energetically the
most stable of all five Bravais lattices.  This is simple to
understand physically, because for different 2D lattices corresponding
to the same density, the triangular lattice is the one with largest
lattice spacing; the electrons thus minimize the repulsive energy by
staying as far away as possible from each other.  Presumably, the
triangular lattice is also the most stable for purely repulsive
potentials of other forms, because the same argument holds.  However,
when one adds an attractive force, one of the other lattice shapes may
become more favorable.  For example, the lattice spacing is a factor
of $(2/\sqrt{3})^{1/2}$ smaller for a square lattice of the same density, and
so the attractive energy can possibly be lowered by choosing this
lattice geometry.  Of course the number of nearest neighbors is also
smaller for the square lattice, so which lattice geometry is
energetically favorable depends on the exact forms and relative
strengths of the attractive and repulsive interactions.

\section{Triangular vs Square Lattice}    \label{section_geometries}

At finite temperature, the lattice geometry that is more favorable is
the one with lower free energy $F = E - TS$.  We will restrict
ourselves to zero temperature, so that it is sufficient to consider
the energy $E$ of the two lattices.  

Using the two-electron potential (eq \ref{potential_total}), one can
simply sum up all pairwise interaction energies for the square lattice
and the triangular lattice, and then compare.  Instead, we will look
at a slightly different quantity, as in BM \cite{bonsall}.  We will
consider the electron located at the origin and let $E$ denote the
energy of the interaction of this electron with all the other
electrons.  The total energy of the lattice of $N$ electrons is then
$\frac{1}{2}NE$.  We compare triangular and square lattices:
\[
E^{\rm TR} = \sum_{\vR_i\neq 0} V(\vR_i^{\rm TR}), \qquad
E^{\rm SQ} = \sum_{\vR_i\neq 0} V(\vR_i^{\rm SQ})
\]
Here $\vR_i^{\rm TR(SQ)}$ runs over the positions of all the electron
positions in the triangular (square) lattice, and $V(\vr)$ is the
potential (eq. \ref{potential_total}) consisting of a Coulomb repulsion
and a surface-mediated attraction.  The two lattices each have the
same density $n$. 

We will look at the difference, 
\be  \label{e_diff}
\Delta E ~=~ E^{\rm TR} - E^{\rm SQ} ~=~ \Delta E_{\rm coul} 
+ \Delta E_{\rm surf}  ,
\ee
between the triangular and square lattices.  The square lattice is
more stable if $\Delta E$ is positive.  The Coulomb part, 
\be  \label{e_diff_coul}
\Delta E_{\rm coul} = 
\sum_{\vR_i\neq 0}^{\rm (TR)} \frac{e^2}{\vR_i^{\rm TR}}
~-~ \sum_{\vR_i\neq 0}^{\rm (SQ)} \frac{e^2}{\vR_i^{\rm SQ}}  ,
\ee
is known to be negative, since the triangular lattice is the most
stable under Coulomb forces alone, and has been studied in detail
in BM \cite{bonsall}.  In terms of density, their
results are
\[
\Delta E_{\rm coul} ~=~  (-3.921034)e^2n^{1/2} - (-3.900265)e^2n^{1/2} 
= -(0.020769)e^2n^{1/2}
\]  
As for the surface part, 
\be \label{e_diff_surf} 
\Delta E_{\rm surf} =
\frac{e^2E_\perp^2}{2\pi\sigma} \lbb -~ \sum_{\vR_i\neq 0}^{\rm (TR)}
K_0 \lba\left|\vR_i^{\rm TR}\right|/l_0\rba ~+~ \sum_{\vR_i\neq
0}^{\rm (SQ)} K_0 \lba\left|\vR_i^{\rm SQ}\right|/l_0\rba \rbb 
= \frac{e^2E_\perp^2}{2\pi\sigma} S_{\rm surf}, 
\ee
it is not {\it a priori} obvious that this is positive, but numerical
calculations confirm that it is.  The transition corresponds to the
values of $E_\perp$ and $n$ for which the two terms just cancel each
other out, $\Delta E_{\rm surf} = - \Delta E_{\rm coul}$, i.e.,
\begin{equation}  \label{transition_curve}
E_\perp^{\rm transition}  ~=~
\sqrt{(2\pi\sigma)(0.020769)(2/\sqrt{3})^{1/2}n^{1/2}  
\lba S_{\rm surf} \rba^{-1} }  .
\end{equation}
To map out the transition line exactly, we need to numerically
calculate $S_{\rm surf}$; the numerical results are shown in figure
\ref{phasediag}.  

We can also analytically predict the dependence of $S_{\rm surf}$ and
$E_\perp$ on the density for larger densities ($n\gg{l}_0^{-2}$).  The
approximation $ K_0 (r/l_0) \approx - \ln(r/l_0)\th(l_0-r)$ is good
for $r\ll l_0$.  Using this ``logarithmic'' approximation, we argue in
appendix \ref{S_surf_append} that $S_{\rm surf}$ becomes independent
of $n$, at large $n$.  Therefore from eq (\ref{transition_curve}) one
finds the shape $E_\perp \sim \sqrt{\sigma}n^{1/4}$ for the transition
curve in the $E_\perp$-$n$ plane, for densities much larger than
$l_0^{-2} = \rho{g}/\s$.

Since the surface-mediated attraction falls off quickly for increasing
inter-electron distances, one would need stronger electric fields at
lower densities to have comparable attractive and repulsive
interactions.  Therefore the transition curve should rise steeply on
the lower-$n$ side, for $n<l_0^{-2}$.  In the logarithmic
approximation the left (lower-$n$) side of the transition curve is
simply a vertical line at $n=l_0^{-2}$.

\section{Zero-Temperature Phase Diagram}   \label{section_0tpd}

\begin{figure}  
\resizebox{12cm}{!}{\includegraphics{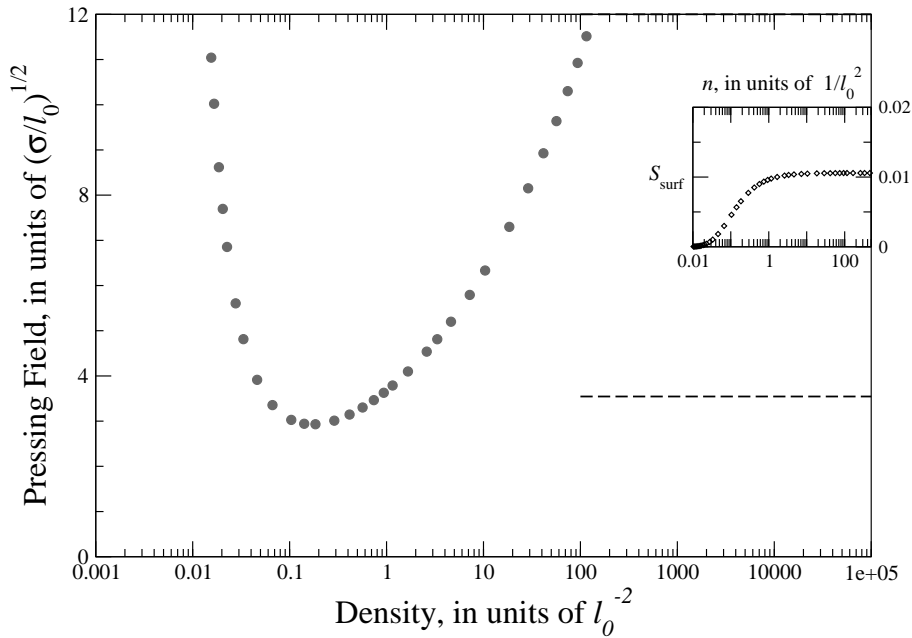}}
\caption{\label{phasediag} $E_\perp$ vs $n$ phase diagram.  Large
dots, from numeric computation, map out the line where triangular and
square lattices have the same energy.  In the region above this curve,
the square lattice is more favorable.  The horizontal line indicates
the spectrum instability, corresponding roughly to the transition to
the surface-buckled, multi-electron dimple state.  At higher densities
this curve should move downward (eq \ref{spectrum_instability}); this
density-dependence is not shown here.  The inset shows the quantity
$S_{\rm surf}$ (eq \ref{e_diff_surf}), determined numerically,
reaching a constant value at densities $\gg{l}_0^{-2}$.
  }
\end{figure}

The zero-temperature phase diagram in the $E_\perp$-density plane is
shown in figure \ref{phasediag}.  $E_\perp$ and $n$ are plotted in
units of $\sqrt{\s/l_0}$ and $1/l_0^2$ respectively; this guarantees
that the same phase diagram is applicable to different cryogenic
substrates, one simply has to replace the numerical values of $\s$ and
$l_0 = \sqrt{\s/g\rho}$ for the particular liquid being used.

The transition line $E_\perp(n)$ is found by calculating $S_{\rm surf}
= -\sum^{\rm TR}K_0(r_i) + \sum^{\rm SQ}K_0(r_i)$ numerically and then
using eq (\ref{transition_curve}).  BM, in doing the corresponding sum
for the Coulomb potential \cite{bonsall}, use an Ewald sum to convert
the summation over $1/r$ to a summation over a faster-decaying
function.  Since $K_0$ is rapidly-decaying itself, no such procedure
is necessary for our case.  Also, BM use a renormalization procedure
(subtracting off contribution due to spread-out positive charge) to
get a finite energy for each lattice geometry.  This is again
unnecessary because the sums converge for the short-ranged $K_0$
function.

Our analytical calculation of the high-density side of the
phase-transition line is actually very good; a $E_\perp \sim n^{1/4}$
fit matches the numeric curve quite spectacularly.  On the lower-$n$
side, the minimum of the curve is at about an order-of-magnitude lower
than $l_0^{-2}$.  The rise of the curve at small density is not
quite as dramatic as the vertical line predicted by the logarithmic
approximation.

At higher densities, there is an $E_\perp$ above which the electrons
cluster into multi-electron dimples (which themselves form a
triangular lattice).  As far as we know, a detailed calculation of the
exact transition line between the single-electron Wigner crystal state
and the surface-buckled multi-electron dimple lattice has never been
performed, but a simple estimate is obtained by considering the
instability in the spectrum of a uniformly charged liquid surface
\cite{ikezi_dimple}:
\begin{equation}  \label{spectrum_instability}
E_\perp = \sqrt{ 4\pi[\rho{g}\s]^{1/2} - (2{\pi}ne)^2 }  \, .
\end{equation}
The low-density part of this curve is the horizontal line in the phase
diagram, figure \ref{phasediag}.  The $n$-dependent deviation due to
the $(2{\pi}ne)^2$ term is substrate-dependent, even with our choice of
units for $E_\perp$ and $n$, and is important only at very high
densities, and so is omitted from the plot.

\section{Concluding Remarks}

In summary, we have reported a novel effect of surface distortions
on the Wigner crystal formed by electrons deposited on a liquid
substrate.  Our analysis shows that, at low electron densities, there
is a significant portion of the zero-temperature $E_\perp$-$n$ phase
diagram where a square lattice is energetically more favorable than
the usual triangular lattice.  Since we have not done a
stability analysis of the square lattice under the combined
interaction of eq \eqref{potential_total}, we cannot yet say whether
the square lattice is stable, or whether some lattice geometry other
than square and triangular is the actual stable geometry.  However,
the energy calculation proves quite clearly the presence of at least
one structural phase transition at low densities.

The structural transition in this system is particularly remarkable
because it arises from the interplay of two very simple forces, a
long-range repulsion and a short-range attraction.  Other physical
systems for which structural transitions have been discussed
(crystalline solids \cite{born_huang}, flux-line lattices, skyrmion
lattices \cite{skyrmion_lattice,skyrmion_geometry}) tend to involve
far more complicated interactions between the constituents.

Experiments on this system have, until now, probed only densities
significantly larger than $l_0^{-2}$.  For $^4$He, the capillary
length corresponds to $l_0^{-2} \approx$ 400 cm$^{-2}$, while typical
experimental Wigner-crystal densities are in the range $\sim
10^5$-$10^9$ cm$^{-2}$.  The same situation holds for explored
electron densities on other surfaces and interfaces.  To the best of
our knowledge, there have been no experimental efforts aimed at
exploring the electron crystal structure at very low densities
($n\sim{l}_0^{-2}$).  One possible experimental signature of a
structural transition to a different geometry would be a shift in the
resonance positions in a Grimes-Adams \cite{grimes} type experiment. 

The analysis in the present work, in addition to the prediction of at
least one structural transition, poses several questions.  First,
there is the issue of the region of parameter space in which the
triangular and square lattices are actually stable.  Stability
questions can be investigated by studying the dynamical matrix or the
elastic constants \cite{born_huang}.  Instability of both lattices in
some part of the $E_\perp$-$n$ plane would indicate that a different
geometry is more favorable than both the lattices we have considered.
(Situations in which more than two lattice types are important have
been encountered previously in the context of skyrmion lattices
\cite{skyrmion_geometry}.)  In such a case there is the added question
of what lattice, or other structure, is the actual stable one.  One
way to find out is to calculate the lattice energies for all five
Bravais lattices (as done for the pure coulomb case in BM
\cite{bonsall}), or better yet, for all possible lattice geometries,
parameterized in a suitable way.  And finally, there is the question
of crossover between single-electron lattice structures to
many-particle dimple lattice structures as one increases the electron
density at high $E_\perp$.  One immediately plausible speculation is
that this transition may proceed via a dimerization.  Several of these
issues are the subject of ongoing calculations and will appear in a
future publication.

\acknowledgments

We thank I. Skachko, R. Chitra, E.Y. Andrei, B.I. Halperin and M.H. Cohen
for useful discussions.

%%%%%%%%%%%%%%%%%%%%%%%%%%%%%%%%%%%%%%%%%%%%%%%%%%%%%%%%%%%%%%%%%%
\appendix

\section{Dependence of $S_{\rm surf}$ on Density}  
\label{S_surf_append}

For densities significantly larger than $n_0 = 1/l_0^2$, we use $
K_0(r/l_0) \approx - \ln(r/l_0)\th(l_0-r)$ and get for the $S_{\rm
surf}$ (defined in eq \ref{e_diff_surf}):
\[
S_{\rm surf} ~\approx~ -~  
\sum_{R_i< l_0 }^{\rm (TR)} 
\ln \lba\left|\vR_i^{\rm TR}\right|/l_0\rba   
~+~ \sum_{R_i<l_0}^{\rm (SQ)}   
\ln \lba\left|\vR_i^{\rm SQ}\right|/l_0\rba   
\]
The summation for each lattice covers all the lattice points
(electrons) within a circular area of radius $l_0$, the number $N =
n(\pi l_0^2)$ of electrons is the same for the two lattices.  The two
terms each contribute a term of magnitude $\pm N\ln(l_0)$, which cancel. 
Rescaling ($r_i = R_i n^{1/2}$), 
\[
S_{\rm surf} ~\approx~ -~  
\sum_{r_i< l_0\sqrt{n}}^{\rm (TR)} \ln \lba r_i^{\rm TR}\rba   
~+~ \sum_{r_i<l_0\sqrt{n}}^{\rm (SQ)}   \ln \lba r_i^{\rm SQ} \rba    
\]
While each sum depends on the radius $l_0n^{1/2}$ of the circle, the
difference does not. This can be seen by considering a radius $r$, and
then increasing the radius by a small amount $\delta{r}$; the number
of electrons in the shell is $\delta{N} = n \times 2\pi{r}\delta{r}$
for either lattice, and therefore the change in each sum is equal to
${\delta}N\log{r}$, which cancel.  There is thus no effect of
increasing the radius $l_0n^{1/2}$ of the circle we're summing over,
i.e., $S_{\rm surf}$ is independent of the density $n$.  A $S_{\rm
surf}$ vs $n$ plot is provided as an inset in figure \ref{phasediag}.

%%%%%%%%%%%%%%%%%%%%%%%%%%%%%%%%%%%%%%%%%%%%%%%%%%%%%%%%%%%%%%%%%%


\begin{thebibliography}{99}
\bibitem{wigner} E.P. Wigner, Phys. Rev. {\bf 46}, 1002 (1934)
\bibitem{eva} E.Y. Andrei, {\it Two-Dimensional Electron Systems
on Helium and other Cryogenic Substrates}, Kluwer Academic Publishers
(1997)
\bibitem{fisher} D.S. Fisher, B.I. Halperin, P.M. Platzman, Phys. Rev.
Lett. {\bf 42}, 798 (1979)
\bibitem{grimes} C.C. Grimes \& G. Adams,  
Phys. Rev. Lett. {\bf 42}, 795 (1979)

\bibitem{shikin_many} V.B. Shikin \& P. Leiderer, Sov. Phys. JETP {\bf
54}(1), 92 (1981). 
\bibitem{leiderer_eva} P. Leiderer, article on the multi-electron
dimple phenomenon, in E.Y. Andrei's compilation \cite{eva}.
\bibitem{gorkov} P. Gor'kov \& D. M. Chernikova,  JETP Lett. {\bf 18},
68 (1974), Sov. Phys. Dokl. {\bf 21}, 328 (1976).
\bibitem{ikezi_dimple} H. Ikezi, R.W. Giannetta, and P.M. Platzman,
Phys. Rev. {\bf B25}, 4488 (1982); also  H. Ikezi,
Phys. Rev. Lett. {\bf 42}, 1688 (1979). 

\bibitem{polaron} The polaronic state for electrons on a liquid
substrate has been discussed by many authors, e.g., N. Studart \&
S. S. Sokolov's article in E.Y. Andrei's compilation \cite{eva}, and
references therein.

\bibitem{bonsall} L. Bonsall and A.A. Maradudin, Phys. Rev. {\bf
B15}, 1959 (1977).

\bibitem{kramer} L. Kramer, Phys. Rev. {\bf B3}, 3821 (1971)
\bibitem{dahm_eva} A.J. Dahm, article discussing electrons on
thin helium films, in E.Y. Andrei's compilation \cite{eva}.

\bibitem{skachko} I. Skachko \& E.Y. Andrei, private communication.

\bibitem{semicond} The possibility exists that the insulating state of
2D electrons in heterojunctions may be some sort of (possibly
modified) Wigner crystal state, see, e.g., S. Chakravarty {\it et al},
Phil. Mag. {\bf B 79}, 859 (1999) and E. Abrahams {\it et al},
Rev. Mod. Phys. {bf 73}, 251 (2001).  In the Quantum Hall community,
there has been discussion of Wigner crystal states interlaced between
Laughlin-liquid states as a function of filling fraction, and related
possibilities, see, e.g., R. Narevich {\it et al}, Phys. Rev. {\bf B
64}, 245326 (2001); K. Yang, F.D.M. Haldane, \& E.H. Rezayi,
Phys. Rev.  {\bf B 64}, 081301 (2001); and references therein.
Also, for an insulating 2D electron system under large magnetic
field, attempts have been made to explain the measured AC response in
terms of an assumed pinned Wigner crystal state, see R. Chitra {\it et
al}, Phys. Rev. {\bf B 65}, (2002), and references therein.
\bibitem{ions} J.N. Tan {\it et al}, Phys. Rev. Lett. {\bf 75}, 4198
(1995); A. W. Vogt, Phys. Rev. Lett. {\bf A 49}, R657 (1994).

\bibitem{skyrmion_lattice} R. C\^ot\'e {\it et al}, Phys. Rev. Lett. {\bf 78},
4825 (1997); and references therein.
\bibitem{skyrmion_geometry} C. Timm, S.M. Girvin, and H.A. Fertig,
Phys. Rev. {\bf B 58}, 10634 (1998); 
S. Sankararaman \& R. Shankar, cond-mat/0209160; 


\bibitem{born_huang} M. Born and K. Huang, \emph{Dynamical Theory of
Crystal Lattices}, Oxford University Press, 1954; reprint 1988.

\bibitem{quantum_computing} A.J. Dahm \emph{et al}, J. Low
Temp. Phys. 126, 709 (2002), quant-ph/0111029;  M.I. Dykman,
P.M. Platzman and P. Seddighrad, cond-mat/0209511.

\end{thebibliography}
\end{document}